\documentclass[onecolumn,sort&compress,numbers]{els-mrw} % For numbered references Style 

\usepackage{amsmath,amssymb,amsfonts,amsthm,makeidx,graphicx}
\usepackage{txfonts}
\usepackage{helvet}\usepackage{mathrsfs}
\usepackage{slashed}

\def\be{\begin{equation}}
\def\ee{\end{equation}}
\def\ba{\begin{array} }

\def\bac{\begin{array} {c}}
\def\bacc{\begin{array} {cc}}
\def\baccc{\begin{array} {ccc}}
\def\bacccc{\begin{array} {cccc}}
\def\ea{\end{array}}
\def\bea{\begin{eqnarray}}
\def\eea{\end{eqnarray}}

\def\Lag{\mathscr{L}}

\newcommand{\beq}{\begin{equation}}
\newcommand{\eeq}{\end{equation}}

\newcommand{\bp}{M_P}

\begin{document}

%%%%%%%%%%%%%%%%%%%%%%%%%%%%%%%%%%%%%%%%%%%%%%%%%%%%%%%%%%%%%%%%
%% the following items are mandatory: 
%% - title
%% - author names
%% - affiliation details
%% - abstract
%% - keywords

%TO DO
%check consistency of notation x
%abstract x
%acronimi x
%conclusions x
%check logic x
%include references with iNSPIRE a salvio x
%English and spelling x
%print and read it all x
%submit to arXiv
%submit to Elsevier

%% Precise, concise, and informative description of the focus of this work. Avoid abbreviations and formulae in the title
\chapter{ Inflationary scenarios beyond the Standard Model}\label{chap1}

%% All author names and affiliations, and email address for corresponding author
\author[1,2]{Alberto Salvio}%
%\author[2]{Second Author}%
%
%\author[1,2]{Third Author}%
 
\address[1]{\orgname{University of Rome Tor Vergata}, \orgdiv{Physics Department}, \orgaddress{via della Ricerca Scientifica, I-00133 Rome, Italy}}
\address[2]{\orgname{I. N. F. N.}, \orgdiv{Rome Tor Vergata}, \orgaddress{via della Ricerca Scientifica, I-00133 Rome, Italy}\\Email: alberto.salvio@roma2.infn.it}

%\articletag{Chapter Article tagline: update of previous edition, reprint.}

\maketitle

%%%%%%%%%%%%%%%%%%%%%%%%%%%%%%%%%%%%%%%%%%%%%%%%%%%%%%%%%%%%%%%%
%% the following item is mandatory: 
%% 100-150 word summary of the chapter
\begin{abstract}[Abstract]
	The aim of this chapter is to explain in clear and pedagogical terms how some particle-physics models and/or mechanisms can naturally lead to inflation and how this can provide testable predictions that can help us find new physics effects. Two well-established features of theoretical particle physics are linked to an essential property of inflation, a naturally-flat inflaton potential: (1) scale invariance, broken by small quantum corrections, and (2) Goldstone's theorem. It is also illustrated how to combine several scenarios of this type to obtain a rather general particle-physics motivated inflationary setup.
\end{abstract}

%% 5-10 words that embody the key topics in the chapter. What terms would someone put into a search engine if they were looking for a chapter like this?
\begin{keywords}
 	Inflation\sep Standard Model \sep beyond the Standard Model \sep cosmology of particle physics models
\end{keywords}

%%%%%%%%%%%%%%%%%%%%%%%%%%%%%%%%%%%%%%%%%%%%%%%%%%%%%%%%%%%%%%%%
%% the following item is optonal: 
%% - Single figure visually illustrating the key topic/method/outcome described in the chapter
%\begin{figure}[h]
%	\centering
%	\includegraphics[width=7cm,height=4cm]{blankfig}
%	\caption{Optional: Single figure visually illustrating the key topic/method/outcome described in the chapter. 
%		     Please add here some text explaining the pic...}
%	\label{fig:titlepage}
%\end{figure}

%%%%%%%%%%%%%%%%%%%%%%%%%%%%%%%%%%%%%%%%%%%%%%%%%%%%%%%%%%%%%%%%
%% the following item is optional: 
%% - System of abbreviations/terms/symbols used in the specific field of study/community. List and define
%\begin{glossary}[Nomenclature]
%	\begin{tabular}{@{}lp{34pc}@{}}
%		LHC & Large Hadron Collider\\
%	\end{tabular}
%\end{glossary}

%%%%%%%%%%%%%%%%%%%%%%%%%%%%%%%%%%%%%%%%%%%%%%%%%%%%%%%%%%%%%%%%
%% the following item is mandatory: 
%% List of the key points and topics a reader can expect to learn from this chapter 
\section*{Objectives}
\begin{itemize}
	\item The general objective of this chapter is learning about the connections between inflation  and particle physics models and/or mechanisms.
	\item First specific goal: understanding the connection between  inflation, scale invariance and its breaking.
	\item Second specific goal: understanding the connection between Goldstone's theorem in particle physics and inflation. 
	\item Third specific goal: combining the above mentioned topics to become familiar with more general scenarios.
\end{itemize}

%%%%%%%%%%%%%%%%%%%%%%%%%%%%%%%%%%%%%%%%%%%%%%%%%%%%%%%%%%%%%%%%
%% the following items are mandatory: 
%% - Section: Introduction 
%% - further sections
%% - Section: Conclusion
\section{Introduction}\label{intro}

(Cosmic) inflation is a period of exponential expansion that took place in the early universe; the expansion one is referring to in giving this definition is the growth in time of the physical distance between two generic distinct space points. This is only possible when the spacetime metric is allowed to change in time, as is the case in the general theory of relativity (GR). 

One of the main motivations for considering inflation is the explanation it provides for the near homogeneity and isotropy of the universe we observe. Indeed, it would be puzzling to explain without inflation why regions of the universe that appear to be causally disconnected have come to a relative thermal equilibrium with a similar temperature. A large-enough exponential expansion, on the other hand, implies that all the observable universe has occupied a much smaller space at an early time.  This allowed all its subregions to be in causal connection.

A pedagogical introduction to GR, cosmology and inflation is out of the scope of the present chapter, because it would require an entire book. Fortunately, there already exist a number of introductory books on these topics. Particularly suitable for the material that is covered here is the approach of the textbooks by Weinberg (1972, 2008), which can be used to find detailed explanations of some results quoted in this chapter concerning GR, cosmology and inflation.  

A reader of an encyclopaedia of particle physics, and in particular of a section on beyond the Standard Model (BSM) may wonder about the importance of inflation in the search for new physics, i.e.~something that is not described by the Standard Model (SM). Typical models of inflation involve energies much larger than those reachable at particle colliders, so the observation of the consequences of inflation can open a useful window on high-energy extensions of the SM. 

This brings us to one of the main successes of inflation, the fact that it provides a quantum-mechanical origin of primordial fluctuations, which have left observable imprints on the cosmic microwave background (CMB), see Weinberg (2008) for an introduction. These imprints have been observed by experiments such as the 
Cosmic Background Explorer (COBE), the Wilkinson Microwave Anisotropy Probe (WMAP) and the observatory Planck of the European Space Agency (ESA). Moreover, they will be further tested by future observations, such as those of LiteBIRD, a satellite that is planned to produce new CMB data during the next decade.

Since the theory of inflation is based on GR and quantum mechanics, it necessarily has to deal with quantum field theory (QFT), the theoretical framework at the basis of particle physics. Therefore, it appears particularly appropriate to include a chapter on inflation on an encyclopaedia that is supposed to provide a comprehensive treatment of elementary particles and their phenomenology. Because of this reason the focus of this chapter is on particle-physics-motivated models, leaving aside other theoretical constructions whose relations with fundamental real-world physics is dubious.

The purpose of this chapter is not to give a pedagogical introduction to the fields of  QFT, inflation and their relation, as these are already explained nicely in a number of textbooks. In particular, the texts of Weinberg (2005, 2008, 2013) provide a complete treatment. The reader can find there detailed explanations of some results quoted in this chapter concerning QFT, inflation and their relation. The aim here is instead to explain in clear and pedagogical terms how some particle-physics models and/or mechanisms can naturally lead to inflation and how this can lead to testable predictions that can help us find new physics effects.
 
 An exponential expansion occurs when the energy density is dominated by a constant term in the potential density. But such an expansion necessarily has to end to avoid an excessive dilution of matter, which would contradict what we observe. A scalar field (or a set of scalar fields) known as the inflaton(s), whose potential feature an approximate plateau for some but not all field values, is then used to trigger inflation. In this context, the quest for new physics is the search of the necessary ingredients to realize such potential flatness in a specific region of field space and to produce the CMB features that we observe in the sky.

%%%%%%%%%%%%%%%%%%%%%%%%%%%%%%%%%%%%%%%%%%%%%%%%%%%%%%%%%%%%%%%%
%% in the following we showcase possible style elements
%%
%%

\section{Elementary inflaton: scale invariance}

One of the consistency requirements of QFT in the spacetime continuum is renormalizability, which allows us to express the predictions of the theory in terms of a finite number of parameters. It is important to keep in mind, however, that this requirement should only be imposed on the microscopic theory. Possible composite-particle states are generically described by non-renormalizable effective field theories. Renormalizable QFTs, on the other hand, enjoy (at least classically) a scale invariance in the high energy limit: indeed, at large-enough energies all dimensionful parameters present in the action can be neglected and we are left only with the marginal interactions, which are dimension-four and thus scale-invariant.

Classical scale invariance is generically broken by quantum effects, which are able to introduce a physical mass scale even in theories with no dimensionful parameters. This effect of radiative generation of scales is known as dimensional transmutation (DT). A notable example of DT is the generation of the confining energy scale of quantum chromodynamics (QCD), $\Lambda_{\rm QCD}$, below which strong interactions cannot be described by perturbation theory and composite states such as the proton, the neutron and the $\pi$ meson form.  A pedagogical introduction to scale symmetry and its breaking by quantum effects is given by Coleman (1985). 

One property of DT is its ability to generate exponentially large hierarchies between mass scales. This is because the renormalization group (RG) running of a generic dimensionless coupling $\kappa$ is logarithmic in the energy scale $\mu$. Therefore, order-one variations of $\kappa$ correspond to exponentially large variations of $\mu$. This is the reason why physicists do not consider the large hierarchy between $\Lambda_{\rm QCD}$ and the Planck mass to be puzzling, but they  rather see it as a quite natural fact.

In Coleman (1973) and Gildener (1976) the radiative generation of scales has been extended to several other theories, different from QCD. These include interesting extensions of the SM, where the Fermi scale (the vacuum expectation value (VEV) of the Higgs field, $v_h$) is radiatively generated. This opened up the possibility of explaining the large hierarchy between $v_h$ and the Planck mass through DT. Furthermore, DT has also been realized in the gravitational context. Salvio (2014, 2018a), Kannike (2015) and Salvio (2021)  showed how to radiatively generate the Planck mass and the cosmological constant (see also Alvarez-Luna (2023) for a subsequent analysis). This is particularly relevant in inflation, where gravitational effects cannot be ignored. These results are pedagogically illustrated in Sec.~\ref{Planckion inflation}.

Before coming to that, it is useful to know that high-energy scale invariance leads  to inflaton potentials that satisfy the important property of quasi flatness for some (but not all) values of the inflaton. Moreover, this scenario leads to almost scale-invariant curvature perturbations, where the spectral index $n_s$  of the curvature power spectrum is approximately $n_s \approx 0.97$, in agreement with observations ($n_s =1$ is the scale invariant case).
These results are illustrated in Secs.~\ref{Higgs(-like) inflation},~\ref{Planckion inflation} and~\ref{Scalaron inflation}. 
High-energy scale invariance is viewed here as a property of elementary inflatons, featuring renormalizable interactions with the other matter fields, as opposed to composite inflatons that are generically  described at low-enough energies by non-renormalizable effective field theories. The latter topic is instead covered in Sec.~\ref{Composite inflaton: the Goldstone case}.

\subsection{Higgs(-like) inflation}\label{Higgs(-like) inflation}

To illustrate how scale invariance can lead to suitable inflationary potentials, let us start with Higgs(-like) inflationary models, namely models featuring an  action for the metric and the inflaton $\phi$ of the form\footnote{Einstein's notation (where repeated indices understand a summation) is used.}
\be S_{g\phi}= \int d^4x\sqrt{-g}\left(\frac{M_P^2+\xi \phi^2}{2}R-\frac12\partial_\mu\phi\partial^\mu\phi -\frac{m^2}{2}\phi^2 -\frac{\lambda}{4}\phi^4-\Lambda_{\rm cc}\right),\label{HiggLike}\ee
where $g$ is the determinant of the metric, $g_{\mu\nu}$, the parameters $M_P$, $\xi$, $m$, $\lambda$ and $\Lambda_{\rm cc}$ are respectively the reduced Planck mass, $M_P\approx 2.4\times 10^{18}$~GeV, a non-minimal coupling between the inflaton $\phi$ and the Ricci scalar\footnote{We take the mostly plus signature convention for the metric and
$$ \Gamma_{\mu \, \nu}^{\,\sigma} \equiv \frac12 g^{\sigma\rho}\left(\partial_\mu g_{\nu\rho}+\partial_\nu g_{\mu\rho}-\partial_\rho g_{\mu\nu}\right), \quad R_{\mu\nu\,\,\, \sigma}^{\quad \rho} \equiv \partial_{\mu} \Gamma_{\nu \, \sigma}^{\,\rho}- \partial_{\nu} \Gamma_{\mu \, \sigma}^{\,\rho} +  \Gamma_{\mu \, \tau}^{\,\rho}\Gamma_{\nu \,\sigma }^{\,\tau}- \Gamma_{\nu \, \tau}^{\,\rho}\Gamma_{\mu \,\sigma }^{\,\tau}, \quad R_{\mu\nu} \equiv R_{\rho\mu\,\,\, \nu}^{\quad \rho},\quad R\equiv g^{\mu\nu}R_{\mu\nu}.$$} $R$, also $m$ is the mass parameter of the inflaton, $\lambda$ its quartic self interaction and $\Lambda_{\rm cc}$ is the cosmological constant. The dimensionful parameters $M_P$ and $m$ can also be generated through DT, as illustrated in Sec.~\ref{Planckion inflation}. In this subsection we are agnostic about their origin. As we will see, the important feature of the action above is that its $\phi$-dependent part features scale invariance at large $\phi$ values. The inflaton $\phi$ can be (but not necessarily has to be) identified with the Higgs field $h$, the real scalar from the Higgs doublet in the unitary gauge, Bezrukov (2008). In that case the $Z_2$-symmetry, $\phi \to-\phi$, of Action~(\ref{HiggLike}) is inherited from the gauge invariance of the SM.

Even if one identifies $\phi=h$, one is not committed to assume that  the full model is just the SM plus Einstein gravity (plus the mentioned non-minimal coupling). It is possible to realize this ``Higgs inflation" in well-motivated SM extensions, which are able to account for  the observational evidence of new physics, see e.g.~Salvio (2015, 2019).

Let us now see how a suitable inflationary potential appears starting from Action~(\ref{HiggLike}). The non-minimal coupling can be eliminated through a field redefinition of the form
\begin{equation} \hat{g}_{\mu \nu}(x)\equiv  \Omega^2(x)  g_{\mu \nu}(x), \label{WeylTr} \ee
where $\Omega^2$ is a positive function of the spacetime point, $x$. The positivity of $\Omega^2$ guarantees that the signature of the metric is preserved. A field redefinition of this sort is also known as a Weyl transformation. In particular, one chooses
\be \Omega^2= 1+\frac{\xi \phi^2}{M_P^2}. \label{transformation}\end{equation}
For $\xi\geq0$ one has $\Omega^2>0$ for all $\phi$. For   $\xi<0$ one has to impose that the relevant values of $\phi$ satisfy $\phi^2<-M_P^2/\xi$ to ensure that $\Omega^2>0$.
After this field redefinition \begin{equation} S_{g\phi} = \int d^4x\sqrt{-\hat{g}}\left[\frac{M_P^2 }{2}\hat{R}-\frac{K}{2}\partial_\mu\phi\hat\partial^\mu\phi-\frac{\frac{m^2}{2}\phi^2 +\frac{\lambda}{4}\phi^4+\Lambda_{\rm cc}}{\Omega^4}\right], \label{EinsteinF}\end{equation}
where $\hat{R}$ is the Ricci scalar computed with the redefined metric $\hat g_{\mu\nu}$, also $\hat\partial^\mu\equiv \hat g^{\mu\nu}\partial_\nu$ and $K\equiv (\Omega^2+6\xi^2\phi^2/M_P^2)/\Omega^4$.
The original (field) frame, where the Lagrangian has the form in (\ref{HiggLike}), is called the Jordan frame, while the one where gravity is canonically normalized (obtained with the Weyl transformation above), Eq.~(\ref{EinsteinF}), is called the Einstein frame. 
The non-canonical Higgs kinetic term in~(\ref{EinsteinF}) can be made canonical through the field redefinition $\phi=\phi(\phi_c)$ defined by
\begin{equation} \frac{d\phi_c}{d\phi}= \sqrt{\frac{\Omega^2+6\xi^2\phi^2/M_P^2}{\Omega^4}}.\label{chi}\end{equation}
Indeed, since the right-hand-side of the expression above is positive, $\phi_c$ is a strictly growing function of $\phi$, which can be inverted to obtain $\phi$ as a function of $\phi_c$.
Thus, the canonically normalized inflaton $\phi_c$ feels a potential 
\begin{equation} \boxed{U\equiv \frac{m^2\phi(\phi_c)^2/2+ \lambda \phi(\phi_c)^4/4+\Lambda_{\rm cc}}{(1+\xi\phi(\phi_c)^2/M_P^2)^2}\label{U}.}\end{equation}
It is now clear that the large-$\phi$ scale invariance of the $\phi$-dependent part of Action~(\ref{HiggLike}) corresponds to a quasi flatness of $U$ at large $\phi$ (or, equivalently, $\phi_c$).

For such high-field values the parameters
\bea \epsilon \equiv\frac{M_P^2}{2} \left(\frac{1}{U}\frac{dU}{d\phi_c}\right)^2, \quad \eta \equiv \frac{M_P^2}{U} \frac{d^2U}{d\phi_c^2}%,\nonumber \\  \zeta^2\equiv \frac{M_P^4}{U^2}\frac{d^3U}{d\phi_c^3} \frac{dU}{d\phi_c} 
\label{epsilon-def}\eea
are thus guaranteed to be small, which ensures that a useful approximation, known as the slow-roll approximation, is valid. In this approximation one can show that $n_s = 1-6\epsilon+2\eta$, computed in a field value corresponding to a large-enough period of inflation, is in agreement with observations for $\phi=h$. Therefore, the same is true also for $\phi\neq h$ if the parameters of the model are chosen close enough to the values corresponding to the SM Higgs case. The ratio $r$ between the curvature and (3D) tensor perturbations (the primordial gravitational waves), computable as $r=16\epsilon$ in the slow-roll approximation, is in the SM Higgs case in agreement with the CMB observations for  a large value of $\xi$, while it turns out to violate a CMB bound, Ade (2021), when $\xi$ is brought down to values of order 10. In Sec.~\ref{A mixed recipe: multifield inflation} it is shown how extending the above-mentioned principle of scale invariance to the pure gravitational action makes Higgs inflation compatible with observations also for an order-10 $\xi$.
To obtain these results the values of the parameters at the inflationary scales are obtained by using the observed values of those parameters in the SM at accessible collider energies and by extrapolating them with the use of the RG. It is important to keep in mind that possible  new physics present in between these two energy scales may or may not change the RG running, altering some of the predictions of Higgs(-like) inflation.

The inflationary model described in this subsection, although it can, does not need to be Higgs inflation. Other inflationary models described by an action of the form~(\ref{HiggLike}) include, for example, Hilltop inflation, Boubekeur (2005), where $|m|$ is taken somewhere in between the electroweak and the Planck scale and $\xi$ can assume negative values, Salvio (2019b), unlike in Higgs inflation.

\subsection{Inflation from radiative generation of scales}\label{Planckion inflation}
This subsection shows how the radiative generation of the scales, $v_h$, $M_P$ and the cosmological constant $\Lambda_{\rm cc}$, offers well-motivated inflaton BSM candidates. 

\subsubsection{Radiative generation of scales}\label{Radiative generation of masses}

To explain this point let us go through the original mechanism of radiative generation of masses proposed by Coleman (1973) and Gildener (1976) without gravity. After that we will see how to take gravity into account. Both with and without gravity the focus here is on a perturbative mechanism of this sort: the requirement of perturbativity allows us to provide an analytical description and, thus, to maintain a pedagogical level.

One assumes that in the tree-level (zero-loop) potential $V$ there are no dimensionful parameters,
  \be V(\varphi) = \frac{\lambda_{\alpha\beta\gamma\delta}}{4!} \varphi_\alpha\varphi_\beta\varphi_\gamma\varphi_\delta, \label{Vns} \ee  
  where the indices $\alpha, \beta, ...$ run over all scalars in the theory. We take the scalar fields $\varphi_\alpha$ real without loss of generality because we allow for a generic number of them. The $\lambda_{\alpha\beta\gamma\delta}$ are the full set of (dimensionless) quartic couplings.
  The mass scales then emerge radiatively from loops in a way we discuss now.

  The basic idea is that, since at quantum level the couplings depend on the energy $\mu$ 
as dictated by the RG, there can be some specific energy %$\tilde\mu$ 
at which the potential  in Eq.~(\ref{Vns}) develops a flat direction. Such direction can be written as $\varphi_\alpha = \nu_\alpha \chi$, where $\nu_\alpha$ are the components of a unit vector $\nu$, i.e.~$\nu_\alpha \nu_\alpha =1$, and $\chi$ is a single scalar field, %such that $\varphi^2 = \phi_a \phi_a$ and 
which parameterizes this direction. 
After renormalization, the potential $V$ along the flat direction, therefore, reads
\be V(\nu \chi) = \frac{\lambda_\chi (\mu)}{4}\chi^4,  \label{Vvarphi}\ee 
where 
\be \lambda_\chi(\mu) \equiv\frac1{3!} \lambda_{\alpha\beta\gamma\delta}(\mu)\nu_\alpha \nu_\beta \nu_\gamma \nu_\delta. \label{lambdaphi}\ee
Having a flat direction along $\nu$ for $\mu$ equal to some specific value $\tilde\mu$ means 
 \be \lambda_\chi(\tilde\mu)\equiv\lambda_{\alpha\beta\gamma\delta}(\tilde\mu)\nu_\alpha \nu_\beta \nu_\gamma \nu_\delta=0. \ee
 The RG running is logarithmic, so starting with, say, an order-one value of 
 $\lambda_\chi(\mu)$ at some high-energy scale $\mu$, the value $\tilde\mu$ at which $\lambda_\chi(\tilde\mu)=0$ is typically exponentially large. Therefore, one can explicitly see that this mechanism can be used to generate large hierarchies in a natural way.
 
 Besides the potential in~(\ref{Vvarphi}),  loop corrections also generate other terms $V_1+V_2+...$, where $V_i$ represents the $i$th-loop correction. 
 The explicit expression of $V_1$ is well known. We can rederive it here by recalling that the effective potential does not depend on $\mu$: the renormalization changes the couplings, the masses and  the fields, but leaves the action invariant. So, we can write
 \be \mu \frac{d}{d\mu}(V+V_1+...) =0.\ee
 Using~(\ref{Vvarphi}), the solution of this equation at the one-loop level reads
 \be V+V_1=\frac{\lambda_\chi (\mu)}{4}\chi^4+ \frac{\beta_{\lambda_\chi}}4\left(\log\frac{\chi}{\mu}+a_s\right)\chi^4,  \ee
 where 
 \be \beta_{\lambda_\chi} \equiv  \mu\frac{d\lambda_{\chi}}{d\mu} \ee
 and  $a_s$ is a renormalization-scheme-dependent quantity. Setting now $\mu=\tilde\mu$, where $\lambda_\chi=0$, one obtains
 \be V_{\rm CW} \equiv V+V_1 = \frac{\bar \beta_{\lambda_\chi}}4\left(\log\frac{\chi}{\chi_0}-\frac14\right)\chi^4,\label{CWpot}\ee
where
 \be \bar\beta_{\lambda_\chi} \equiv \left[\beta_{\lambda_\chi} \right]_{\mu=\tilde\mu}, \qquad \chi_0\equiv \frac{\tilde\mu}{e^{1/4+a_s}}.\ee
$V_{\rm CW}$ is known as the Coleman-Weinberg potential. We see that the flat direction gets some steepness at loop level.  The field value $\chi_0$ is a stationary point of $V_{\rm CW}$.
Moreover, $\chi_0$ is a point of minimum when $\bar\beta_{\lambda_\chi}>0$. Therefore, when the conditions 
  \beq\left\{
\begin{array}{rcll}
\lambda_\chi(\tilde\mu)  &=& 0 & \hbox{(flat direction),}\\

\beta_{\lambda_\chi}(\tilde\mu)  &>& 0 & \hbox{(minimum condition),}
\end{array}\right.
\label{eq:CWgen}
\eeq
 are satisfied quantum corrections generate a minimum of the potential at a non-vanishing value of $\chi$, i.~e.~$\chi_0$. In that case $\chi_0$ is the (radiatively induced)  VEV of $\chi$. 
 
  This non-trivial minimum can generically break global and/or local symmetries and thus generate the  particle masses. Consider for example a term in the Lagrangian density 
  $\mathscr{L}$ of the form 
 \be \mathscr{L}_{\chi h}\equiv \frac12 \lambda_{\alpha\beta} \varphi_\alpha\varphi_\beta |H|^2,\label{LvarphiH}\ee 
 where $H$ is the SM Higgs doublet and the $\lambda_{\alpha\beta}$ are some of the quartic couplings. Substituting the coefficients with the corresponding running quantities and setting $\mu=\tilde\mu$ and $\varphi$ along $\nu$, 
 \be  \mathscr{L}_{\chi H} = \frac12 \lambda_{\chi h}(\tilde\mu) \chi^2 |H|^2, \ee
 where 
 \be \lambda_{\chi h}(\mu) \equiv  \lambda_{\alpha\beta}(\mu) \nu_\alpha\nu_\beta.\ee
 Thus, by evaluating this term at the minimum $\chi=\chi_0$ we obtain the Higgs squared mass parameter
  \be \mu_h^2 = \frac12\lambda _{\chi h}(\tilde\mu) \chi_0^2. \ee
 In order to trigger the electroweak symmetry breaking  through the usual Higgs mechanism, we need $\mu_h^2>0$, namely we have the additional condition
 \be \lambda _{\chi h}(\tilde\mu) >0. \ee
 
 Let us move now to the gravitational case. The crucial difference with respect to the non-gravitational case is that the Lagrangian\footnote{In the presence of gravity the full Lagrangian density is $\sqrt{-g}\mathscr{L}$, which is related to the action $S$ through $S=\int d^4x\sqrt{-g}\mathscr{L}$.} can include another $\varphi$-dependent term, involving non-minimal couplings $\xi_{\alpha\beta}$: 
 \be \mathscr{L}_{\rm non-minimal} = \frac12 \mathscr{F}(\varphi)  R, \qquad \mathscr{F}(\varphi) \equiv \xi_{\alpha\beta} \varphi_\alpha\varphi_\beta. \label{Lnonminimal}\ee
  Let us consider again a direction in field space $\varphi_\alpha = n_\alpha \phi$, with $n_\alpha$ the components of a unit vector $n$, that is $n_\alpha n_\alpha =1$, generically different from $\nu$, and $\phi$ parameterizing this direction (the symbol $\phi$ is used again because in Subsubsec.~\ref{DT inflation} this field will be identified with the inflaton, like in Subsec.~\ref{Higgs(-like) inflation}).  Now, define 
 \be \lambda_\phi(\phi) \equiv \frac1{3!}\lambda_{\alpha\beta\gamma\delta}(\phi)n_\alpha n_\beta n_\gamma n_\delta, \qquad \xi_\phi(\phi) \equiv \xi_{\alpha\beta}(\phi)n_\alpha n_\beta, \label{lambdaxiphi}\ee
 where we have set $\mu = \phi$ because our focus here is on the field direction $n$. 
With these definitions the field equation for a homogeneous scalar $\phi$ reads
\be \frac{dV}{d\phi} = \frac{R}{2}\frac{d}{d\phi}\left(\xi_\phi  \phi^2\right) , \label{varphiEOM} \ee
where the dependence of $V$ on $\phi$ is given by
\be V(n \phi) = \frac{\lambda_\phi (\phi)}{4}\phi^4.  \label{Vvarphi2}\ee 
On the other hand, the trace of the gravitational field equations gives
\be 4 V-\xi_\phi(\phi) \phi^2 R = \mathcal{O}(R^2), \label{traceEEOM}\ee
where $\mathcal{O}(R^2)$ represents contributions due to possible quadratic-in-curvature terms, which respect classical scale invariance. A dedicated discussion of such terms is given in Subsec.~\ref{Scalaron inflation} and Sec.~\ref{A mixed recipe: multifield inflation}.
 Now, we want to generate the Planck scale and the cosmological constant radiatively. The observed value of the cosmological constant is tiny and we can therefore  neglect the term $\mathcal{O}(R^2)$. Then, by solving Eq.~(\ref{traceEEOM}) for $R$ and plugging the result in Eq.~(\ref{varphiEOM}), one finds
\be  \frac{dV}{d\phi}=\frac{2V}{\xi_\phi \phi^2}\frac{d}{d\phi}\left(\xi_\phi  \phi^2\right). \ee 
Using~(\ref{Vvarphi2})
 this equation can be rewritten as
\be \beta_{\lambda_\phi} = 2\frac{\lambda_\phi}{\xi_\phi} \beta_{\xi_\phi}, \label{agrGen} \ee 
where 
\be \beta_{\lambda_\phi}\equiv  \phi\frac{d\lambda_\phi}{d\phi}, \qquad \beta_{\xi_\phi}\equiv  \phi\frac{d\xi_\phi}{d\phi}.  \ee 

When Eq.~(\ref{agrGen}) is satisfied at some non-vanishing field value $\bar\phi$  the Planck scale and the cosmological constant $\Lambda_{\rm cc}$ are generated. Specifically, we have 
\be \bp^2=\xi_\phi(\bar\phi) \bar\phi^2, \qquad  \Lambda_{\rm cc}= \lambda_\phi(\bar\phi) \bar\phi^4. \label{MLgen}\ee
 Requiring $\bp^2$ to be positive and  $\Lambda_{\rm cc}$ to match the  observed value (which is negligibly small compared to $\bp^4$)   one obtains the three conditions 
\beq\left\{
\begin{array}{rcll}
\xi_\phi(\bar\phi)  &>& 0 & \hbox{(real Planck mass).}\\
\lambda_\phi(\bar\phi) &=& 0 & \hbox{(nearly vanishing cosmological constant),}\\
\beta_{\lambda_\phi}(\bar\phi) &=& 0 & \hbox{(solution of field  equations),}
\end{array}\right.
\label{eq:agravMPl}
\eeq
where Eq.~(\ref{agrGen}) has been simplified taking into account that $\lambda_\phi$ nearly vanishes at $\bar\phi$. These are the necessary and sufficient conditions to generate perturbatively  through DT viable values of $M_P$ and $\Lambda_{\rm cc}$. 
%Note that the specific value of $\bar\varphi$ is not relevant because $\bar\varphi$ is the only source of mass scale and therefore  can  be set to 1 by choosing the unit of mass appropriately. 
We will refer to the scalar field $\phi$ defined in this subsection as ``the Planckion" as it is responsible for $M_P$. 
%Note that the CW mechanism (through the scalar $\chi$) can also occur in the presence of gravity, but that is not sufficient to have a viable scenario: if the CW mechanism were the only source of scales, then the cosmological constant would be negative~\cite{Salvio:2014soa,Elizalde:1994gv}: setting $\chi=\chi_0$ in~(\ref{CWpot}) we obtain a negative value.

It is interesting to compare the conditions for successful radiative generation of masses without gravity, Conditions~(\ref{eq:CWgen}), and with gravity, Conditions~(\ref{eq:agravMPl}). They are significantly different. The only conditions that are superficially similar are the first one in~(\ref{eq:CWgen}) and the second one in~(\ref{eq:agravMPl}). Note, however, that they have a completely different origin. While the former is the requirement of having a flat direction, the latter corresponds to having a negligibly small value of $\Lambda_{\rm cc}$ in the presence of gravity.

Finally, note that the particle masses in the matter sector can be generated through the quartic couplings that couple $\phi$ and/or $\chi$ to the Higgs fields of the theory. The couplings of $\chi$ to the Higgs fields have already been discussed. Similarly, for $\phi$, if there is a term of the form~(\ref{LvarphiH}) one obtains the Higgs squared mass parameter 
 \be \mu_h^2 = \frac12\lambda _{\phi h}(\bar\phi) \bar\phi^2, \ee
 where
  \be \lambda_{\phi h}(\phi) \equiv  \lambda_{\alpha\beta}(\phi) n_\alpha n_\beta.\ee
 and the condition to generate the particle masses through the usual Higgs mechanism is 
 \be \lambda _{\phi h}(\bar\phi) >0. \ee
% In the general case both $\chi_0$ and $\bar\phi$ can contribute to $\mu_h^2$: 
%  \be \mu_h^2 = \frac12\lambda _{\phi h}(\bar\phi) \bar\phi^2 +\frac12\lambda _{\chi h}(\tilde\mu) \chi_0^2 \ee

%English and spelling up to here 

\subsubsection{Inflation}\label{DT inflation}

Let us now move to inflation in theories with radiative generation of masses. A priori we have two inflaton candidates, $\chi$ and $\phi$. It has been pointed out, however, that identifying the inflaton with $\chi$ leads to a prediction for $n_s$ in disagreement with CMB observations, at least in the absence of non-minimal couplings, Barenboim (2014). Let us then focus on $\phi$, i.e.~``Planckion inflation". 

Performing the field redefinition in~(\ref{WeylTr}) with $\Omega^2 = \mathscr{F}(\varphi)/M_P^2$ one eliminates the non-minimal coupling and obtains the following ``Einstein-frame" potential of $\phi$: 
\be U \equiv \bp^4 \frac{V}{\mathscr{F}^2}. \ee
Substituting $V$ with $\lambda_\phi \phi^4/4$ and  $\mathscr{F}$ with $\xi_\phi \phi^2$ (see Eq.~(\ref{lambdaxiphi})), one finds
\be \boxed{U = \frac{\bp^4}{4}\frac{\lambda_\phi}{\xi_\phi^2}.} \label{UPlanckion}\ee 
The potential that $\phi$ feels is, therefore,  flat at tree-level. The same radiative corrections that lead to DT also lead to some slope, which, however, is small because we have assumed the theory to be perturbative. One can canonically normalize $\phi$ with the same field redefinition used for Higgs(-like) inflation, Eq.~(\ref{chi}), with the substitution $\xi\to\xi_\phi$.

In Planckion inflation the slow-roll parameters in~(\ref{epsilon-def}) are, therefore,  given by the beta functions of $\lambda_\varphi$, $\xi_\varphi$ (and their beta functions as $\eta$ in~(\ref{epsilon-def}) involves the second derivative of the potential). Explicit expressions for $\epsilon$ and $\eta$ can be found in Salvio (2014). So, the slow-roll parameters are  automatically suppressed by loop factors as the theory is assumed to be perturbative in this case; no ad hoc assumptions are needed to make them small and to have a plateau in the inflationary potential.

The exact form of the potential is  model dependent as it depends on the beta functions. However, if we expand $U$ around the field value $\bar \phi$ introduced in Subsubsec.~\ref{Radiative generation of masses} and we stop at the quadratic order in $\phi- \bar\phi$ we have
\be U \approx \frac{M_\phi^2}{2} (\phi - \bar\phi)^2, \label{Uplanckionapp}\ee
where we have used 
\be \phi \frac{\partial U}{\partial\phi} = \frac{\bp^4}{4}\left(\frac{\beta_{\lambda_\phi}}{\xi_\phi^2} -\frac{2\lambda_\phi}{\xi_\phi^3} \beta_{\xi_\phi}\right), \ee
which follows from~(\ref{UPlanckion}),
and the fact that both $\lambda_\phi$ and $\beta_{\lambda_\phi}$  vanish at $\bar\phi$ (see~(\ref{eq:agravMPl})). The quantity $M_\phi^2$ in~(\ref{Uplanckionapp}) is a non-negative constant, see Salvio (2021), defined by
\be M_\phi^2 \equiv \frac{d^2U}{d\phi^2}(\bar\phi), \ee which we interpret as the Planckion squared mass. So in this approximation the form of the Planckion potential is universally quadratic in any model.
Using also the slow-roll approximation, one can show that $n_s = 1-6\epsilon+2\eta$, computed in a field value corresponding to a large-enough period of inflation, is in agreement with observations in Planckion inflation.  The ratio $r$, however, violates a CMB bound, Ade (2021). As for some Higgs(-like) inflationary scenarios, in Sec.~\ref{A mixed recipe: multifield inflation} it is shown how including quadratic-in-curvature terms in the action makes also Planckion inflation fully compatible with observations.

\subsection{Generalizations and scalaron inflation}\label{Scalaron inflation}

It is interesting to note that adding quadratic-in-curvature terms to $\mathscr{L}$,  
\be  \mathscr{L}_{\rm quad} = \alpha R^2  +\beta R_{\mu\nu}R^{\mu\nu}+\gamma R_{\mu\nu\rho\sigma}R^{\mu\nu\rho\sigma} 
 \label{Lquad}
 \ee 
is natural in the context of scale invariance because they also do not involve any mass parameter, i.e.~the coefficients $\alpha$, $\beta$ and $\gamma$ are dimensionless.

One combination of the terms in (\ref{Lquad})  is a total (covariant) derivative, the ``topological Gauss-Bonnet term":
 \beq   G\equiv  R_{\mu\nu\rho\sigma}R^{\mu\nu\rho\sigma} - 4 R_{\mu\nu} R^{\mu\nu} + R^2  =  \frac{1}{4}\epsilon^{\mu\nu\rho\sigma}
\epsilon_{\alpha\beta\gamma\delta}R_{\,\,\,\,\,\, \mu\nu}^{\alpha\beta} R_{\,\,\,\,\,\,\rho\sigma}^{\gamma\delta}= \mbox{divs.}, \label{Gdef}
 \eeq
where $\epsilon_{\mu\nu\rho\sigma}$ is the antisymmetric Levi-Civita tensor and ``divs" represents the covariant divergence of some current.  This total derivative does not contribute to the field equations and can be often  ignored. Therefore, it is  convenient to write 
(\ref{Lquad}) as 
 \be  \mathscr{L}_{\rm quad} = (\alpha - \gamma) R^2  +(\beta +4\gamma) R_{\mu\nu}R^{\mu\nu}+\gamma G. \label{Lquad2}
 \ee 
 Furthermore, for reasons that will become apparent  soon, it is also convenient to express $R_{\mu\nu}R^{\mu\nu}$ in terms of $W^2\equiv W_{\mu\nu\rho\sigma}W^{\mu\nu\rho\sigma}$, where $W_{\mu\nu\rho\sigma}$ is the Weyl tensor
\beq W_{\mu\nu\alpha\beta} \equiv R_{\mu\nu\alpha\beta} + \frac{1}{2} (
g_{\mu\beta} R_{\nu\alpha} -g_{\mu\alpha} R_{\nu\beta}+ g_{\nu\alpha} R_{\mu\beta} - g_{\nu\beta}R_{\mu\alpha})
+\frac16 (g_{\mu\alpha}g_{\nu\beta}-g_{\nu\alpha}g_{\mu\beta} )R.
\eeq
One has
\beq  \frac12 W_{\mu\nu\rho\sigma}W^{\mu\nu\rho\sigma}  =  \frac12 R_{\mu\nu\rho\sigma}R^{\mu\nu\rho\sigma} - R_{\mu\nu}R^{\mu\nu} + \frac16 R^2, \eeq
which, together with the definition of $G$ in (\ref{Gdef}), gives
\be  R_{\mu\nu}R^{\mu\nu}   =  \frac{W^2}{2} + \frac{R^2}{3} - \frac{G}{2}. \label{RicciSquared} \ee 
By inserting this expression of $R_{\mu\nu}R^{\mu\nu}$ in (\ref{Lquad2}) one finds 
 \be     \mathscr{L}_{\rm quad} =  \frac{R^2}{6f_0^2}   - \frac{W^2}{2 f_2^2}-\epsilon_1 G, \label{Lquad3}
 \ee 
 where 
 \be f_0^2 \equiv  \frac1{2\beta +2\gamma+6\alpha}, \quad f_2^2 \equiv -\frac{1}{\beta + 4\gamma}, \quad \epsilon_1 \equiv \frac{\beta}{2}+\gamma.\ee

These quadratic-in-curvature terms, in particular $R^2$, were used to engineer the first model of inflation, Starobinsky (1980), which was published even before the physics community became interested in inflation. Let us now discuss how $R^2$ can lead to inflation.

It is not necessary (although it is possible) to assume that the dimensionful quantities are radiatively generated with the mechanisms described in Subsubsec.~\ref{Radiative generation of masses}. So, for the sake of generality, let us start by considering an arbitrary theory containing up to dimension-four operators: 
\be \mathscr{L}= \mathscr{L}_{\rm quad} +\frac{\bp^2}{2} R-\epsilon_2 \Box R +\mathscr{L}_{\rm non-minimal}+\mathscr{L}_{\rm matter},\ee
where we have allowed for an extra ``divs" term, with an extra coefficient $\epsilon_2$, and $\mathscr{L}_{\rm non-minimal}$ is again given by~(\ref{Lnonminimal}).
The general matter content  includes real scalars $\varphi_\alpha$, Weyl fermions $\psi_j$ and vectors $V^A_\mu$ (with field strength $F_{\mu\nu}^A$) and its Lagrangian is given by 
\bea\Lag_{\rm matter} &=&  
- \frac14 F_{\mu\nu}^AF^{\mu\nu A} - \frac{D_\mu \varphi_\alpha \, D^\mu \varphi_\alpha}{2}  + \bar\psi_j i\slashed{D} \psi_j  - \frac12 (Y^\alpha_{ij} \psi_i\psi_j \varphi_\alpha + \hbox{h.c.}) \nonumber\\
&& 
- V(\varphi) -\frac12 (M_{ij}\psi_i\psi_j+\hbox{h.c.}), \label{Lmatter}
\eea %
where 
\be V(\varphi) = \frac{m^2_{\alpha\beta}}{2} \varphi_\alpha \varphi_\beta + \frac{A_{\alpha\beta\gamma}}{3!}\varphi_\alpha\varphi_\beta\varphi_\gamma+ \frac{\lambda_{\alpha\beta\gamma\delta}}{4!} \varphi_\alpha\varphi_\beta\varphi_\gamma\varphi_\delta+\Lambda_{\rm cc}, \label{RenPot} \ee
%\xxx{referee2:} {\color{blue} 
and all terms are contracted in a gauge-invariant way.
The covariant derivatives are\footnote{The spin-connection $ \omega^{ab}_\mu$ is defined as usual by 
$ \omega^{ab}_\mu = e^a_{\,\, \nu} \partial_\mu e^{b\nu} + e^a_{\,\, \rho}  \Gamma_{\mu \, \sigma}^{\,\rho}  e^{b\sigma}$, where the vierbein $e^a_{\,\, \mu}$ satisfies by definition $e^a_{\,\, \mu}e^b_{\,\, \nu}\eta_{ab}=g_{\mu\nu}$,
and $\gamma_{ab} \equiv \frac14 [\gamma_a, \gamma_b]$. Here $\eta_{ab}$ and $\gamma_a$ are the metric and Dirac matrices on Minkowski spacetime. The spin connection guarantees that the action is invariant not only under general coordinate transformations, but also under   $e^{a}_\mu \to \Lambda^a_{~b}e^b_\mu$, where $\Lambda^a_{~b}$  
are the elements of a local (spacetime dependent) Lorentz matrix. This gauge symmetry is necessary as it leaves the defining property of the vierbein  $e^a_{\,\, \mu}e^b_{\,\, \nu}\eta_{ab}=g_{\mu\nu}$ invariant.} 
$$D_\mu \varphi_\alpha = \partial_\mu \varphi_\alpha+ i \theta^A_{\alpha\beta} V^A_\mu \varphi_\beta \qquad D_\mu\psi_j = \partial_\mu \psi_j + i t^A_{jk}V^A_\mu\psi_k + \frac12 \omega^{ab}_\mu \gamma_{ab}  \psi_j.$$ 
The gauge couplings are contained in the matrices $\theta^A$ and $t^A$, which are the generators of the gauge group in the scalar and fermion representation respectively, while $Y^\alpha_{ij}$ and $\lambda_{\alpha\beta\gamma\delta}$ are the Yukawa  and  quartic couplings respectively. We have also added general renormalizable mass terms and cubic scalar interactions.
 Of course, for specific assignments of the gauge and global symmetries some of these parameters can vanish, but here we  keep a general expression.  When the scales are radiatively generated we can simply take all dimensionful coefficients in $\mathscr{L}$ to zero. Note  that this theory includes a vast set of BSM scenarios of physical interest.
 
  The non-standard $R^2$ term can be removed by adding to the Lagrangian the term
\be  - \sqrt{-g} \,   \frac{(R-3f_0^2 \Xi/2)^2}{6f_0^2},\ee
where  $\Xi$ is an auxiliary field. This Lagrangian vanishes once the $\Xi$  field equations are used and
we are, therefore, free to add it to the total Lagrangian. However, this has the effect of modifying the non-minimal couplings: the term linear in $R$ in the Lagrangian now reads 
\be  \frac12 \sqrt{-g} \,  F(\Xi,\varphi) R , \qquad  \quad F(\Xi,\varphi)\equiv\bp^2 + \mathscr{F}(\varphi)+\Xi.\ee
In order to get rid of this remaining non-standard   term we perform a ``Weyl transformation": 
\be g_{\mu\nu} \rightarrow \frac{\bp^2}{F}g_{\mu\nu} , \qquad \varphi_\alpha \rightarrow \left(\frac{F}{\bp^2}\right)^{1/2} \varphi_\alpha, \qquad  \psi_j \rightarrow \left(\frac{F}{\bp^2}\right)^{3/4}  \psi_j, \qquad V_\mu^A \rightarrow  V_\mu^A. \label{ConfTransf}\ee
After this transformation the Lagrangian density reads $\sqrt{-g}\mathscr{L}_E$ with 
 \bea    \mathscr{L}_E &=&   - \frac{W^2}{2 f_2^2}+\frac{\bp^2}{2} R + \mbox{divs.}- \frac14 F_{\mu\nu}^AF^{\mu\nu A} + \bar\psi_j i\slashed{D} \psi_j  - \frac12 (Y^\alpha_{ij} \psi_i\psi_j \varphi_\alpha + \hbox{h.c.})-\frac{\sqrt{6}\bp}{2 \zeta}  (M_{ij}\psi_i\psi_j+\hbox{h.c.}) \nonumber\\
&&-\frac{6\bp^2}{\zeta^2} \, \frac{D_\mu \varphi_\alpha \, D^\mu \varphi_\alpha  + \partial_\mu\,  \zeta \partial^\mu\zeta}{2} -  U(\zeta,\varphi), \label{LE}
\eea
where we defined\footnote{Notice that in order for the metric redefinition in (\ref{ConfTransf}) to be regular one has to have $F>0$ and thus we can safely take the square root of $F$.} $\zeta\equiv\sqrt{6F}$ and
\be \boxed{U(\zeta,\varphi) \equiv \frac{36 \bp^4}{\zeta^4}\left[V(\varphi) +\frac{3f_0^2}{8} \left(\frac{\zeta^2}{6} - \bp^2 - \mathscr{F}(\varphi)\right)^2\right].} \label{Udef}\ee
In $ \mathscr{L}_E$ we have not written explicitly the total derivatives as they typically do not play an important role in cosmology.
These total derivatives emerge when the Weyl transformation is applied to the  terms proportional to $\epsilon_1$ and $\epsilon_2$. 

The advantage of this form of the Lagrangian, known as the ``Einstein frame", is the absence of the non-minimal couplings and  the $R^2$ term. The latter has effectively been traded with the new scalar $\zeta$, ``the scalaron" which appears non-polynomially: the scalar kinetic terms are non-canonical and cannot be put in the canonical form with further field redefinitions given that the scalar-field metric, $\frac{6\bp^2}{\zeta^2} I_s$ with $I_s$ the identity matrix in the scalar-field vector space,  is not flat in the presence of other scalar fields besides $\zeta$; moreover, the Einstein frame potential $U$ differs considerably from the original one, $V$. 
Notice that the $W^2$ term is  unchanged in the Einstein frame, this is the reason why $R_{\mu\nu}R^{\mu\nu}$ has been  expressed in terms of  $W^2$, $R^2$ and $G$, with~(\ref{RicciSquared}).
 
Now, ``scalaron inflation" corresponds to identifying the inflaton  with $\zeta$. This occurs when the $\zeta$ direction in the scalar-field space is flatter than the other field directions. In this case the $\varphi_\alpha$ are approximately confined at the minimum of the potential and are only allowed to do small fluctuations.  In this case one can make the kinetic term of $\zeta$ canonical through the field  redefinition $\zeta =\sqrt{6}\bp \exp(\omega/(\sqrt{6}\bp))$. The new field $\omega$, the canonically-normalized scalaron, feels a potential 
\be \boxed{U(\omega) =  \Lambda_{\rm cc} e^{-4 \omega/\sqrt{6}\bp} + \frac{3f_0^2 \bp^4}{8}\left(1-e^{-2\omega/\sqrt{6}\bp}\right)^2,}\ee
where we have neglected $\mathscr{F}(\varphi)$ and all field-dependent terms in $V(\varphi)$ as they can be absorbed in $\bp^2$ and $\Lambda_{\rm cc}$ when the scalar fields $\varphi_\alpha$ are at the minimum of the potential.  There is a stationary point of $U$ for
\be e^{-2\omega/\sqrt{6}\bp} = \frac{3f_0^2 \bp^4/8}{\Lambda_{\rm cc} + 3f_0^2 \bp^4/8}, \ee
whenever the right-hand side of the equation above is positive. For positive cosmological constant, $\Lambda_{\rm cc}>0$, such stationary point always exists for $f_0^2 > 0$ when it is a point of minimum, but for $f_0^2<0$ either the stationary point does not exist or it is a point of maximum, not of minimum. This is a special case of a more general result (which is also valid when the other scalars $\varphi_\alpha$ can fluctuate freely), which   proves that a minimum of the potential exists only for $f_0^2 > 0$, Salvio (2018b). In other words, $f_0^2 > 0$ is necessary to avoid tachyonic instabilities, and thus is assumed here. The larger $\omega/M_P$ is the flatter the scalaron potential $U(\omega)$ turns out to be. This supports slow-roll inflation with all predictions, including $n_s$ and $r$, in perfect agreement with CMB bounds. 

It is worth noting that the Einstein frame Lagrangian, Eq.~(\ref{LE}), not only includes scalaron inflation, but also Higgs(-like) inflation and Planckion inflation, as particular cases. Generally speaking, we can have inflation along a given direction in  field space if the other directions are steeper. This can be achieved by taking appropriate limits of the parameters. This point is further discussed in Sec.~\ref{A mixed recipe: multifield inflation}.

\section{Composite inflaton: the Goldstone case}\label{Composite inflaton: the Goldstone case}

Besides the high-energy scale invariance mentioned above there is another particle-physics mechanism that naturally generates a suitable inflationary potential: identifying the inflaton $\phi$ with a pseudo-Nambu-Goldstone boson (PNGB). Indeed, Goldstone's theorem ensures a flat potential and explicit symmetry breaking terms introduce some steepness, which is necessary for inflation to end.

In this chapter we focus on the simplest particle-physics-inspired possibility: let us consider a version of QCD with a confinement scale $f$ (analogous to, but taken at a much higher value than, the confining energy scale of QCD)  and with three flavors of tilde-quarks: $\tilde q=\{\tilde u, \tilde d, \tilde s\}\equiv \{\tilde q_1, \tilde q_2, \tilde q_3\}$, where the tilde distinguishes from the analogous QCD quantities.
 Like in ordinary QCD the strong dynamics forms condensates with a typical scale  $\tau$: as discussed in the textbook by Weinberg (2008)
\be \langle  \bar{\tilde q}'_i \tilde q'_j \rangle = -\tau \delta_{ij}, \qquad \langle\bar{\tilde q}'_i \gamma_5\tilde q'_j \rangle =0, \label{qVEV} \ee  
where $\langle\cdot\rangle$ represents the vacuum expectation value and $\tilde q'_i$  are the Goldstone-free quark fields:
\be \tilde q' = \exp(i \gamma_5 B/(\sqrt{2}f)) \tilde q. \label{qp}\ee 
The scale $f$ is also analogous to the pion decay constant and  $B$ is the Hermitian matrix containing the tilde-mesons (which have canonically normalized kinetic terms), Weinberg (2008),
\be \label{Bf}   B\equiv \left(\baccc \frac{\tilde\pi^{0}}{\sqrt{2}} +\frac{\tilde\eta^0}{\sqrt{6}}& \tilde\pi^+ & \tilde K^+ \\
(\tilde\pi^+)^\dagger & -\frac{\tilde\pi^{0}}{\sqrt{2}} +\frac{\tilde\eta^0}{\sqrt{6}} & \tilde K^0 \\
(\tilde K^+)^\dagger & (\tilde K^0)^\dagger & -\sqrt{\frac{2}{3}}\tilde\eta^0 \ea \right). 
\ee
These scalars are the Goldstone bosons associated with the breaking of the axial part of the global ${\rm SU(3)}_{\rm f}$  flavor group (which, by definition, rotates $\{\tilde u, \tilde d, \tilde s\}$). Just like in ordinary QCD, one can add quark mass terms  that explicitly break the axial part of ${\rm SU(3)}_{\rm f}$:
\be \Lag_{\rm mass}= -\bar{\tilde q} M_q\tilde q = -\bar{\tilde q}'  \exp(-i \gamma_5 B/(\sqrt{2}f)) M_q \exp(-i \gamma_5 B/(\sqrt{2}f)) \tilde q', \ee
where $M_q$ is the tilde-quark mass matrix. The group ${\rm SU(3)}_{\rm f}$ is an approximate symmetry as long as the elements of $M_q$ are small compared to the other mass scale, $f$.
The tilde-meson potential can be computed  from these mass terms using~(\ref{qVEV}): 
in general this potential turns out to be equal to \be V = \tau\,  {\rm Tr}\left[ \cos\left(\sqrt{2}B/f\right) M_q\right] +\Lambda_0,\label{VNq}\ee
where $\Lambda_0$ is a ``bare"  cosmological constant term.

Using standard effective field theory methods (see again the textbook by Weinberg (2008)) and taking $M_q$ diagonal for simplicity,
\be M_q = {\rm diag}(m_{\tilde u},m_{\tilde d},m_{\tilde s}), \ee
one finds the following spectrum of the eight real PNGBs
\bea  m^2_{\tilde K^0} &=& \frac{\tau}{f^2}(m_{\tilde d} + m_{\tilde s}), \label{K0} \\ 
m^2_{\tilde K^+} &=& \frac{\tau}{f^2}(m_{\tilde u} + m_{\tilde s}), \label{Kp} \\ 
m^2_{\tilde \pi^0} &=& m^2_{\tilde \pi^+}  = \frac{\tau}{f^2}(m_{\tilde u} + m_{\tilde d}),  \label{pi0}\\
m^2_{\tilde \eta^0} &=& \frac{\tau}{f^2}\left(\frac{m_{\tilde u}+m_{\tilde d} + 4 m_{\tilde s}}{3}\right), \label{eta0}
\eea
%(See Sec. 19.7 of~\cite{Weinberg2} )
By using the known values of the meson masses, the up and down quark masses and the pion decay constant one finds
% \xxx{I used that $\tau/f^3$ is not changed much by non-vanishing quark masses. This should be correct because $F_\pi \leftrightarrow f$ and $v\leftrightarrow \tau}$ (see  Sec. 19.7 of~\cite{Weinberg2}) are much bigger than the the up and down quark masses (which generate the pion mass) and the relation between $m^2_{\pi^0}$ and $m_u$ and $m_d$ is the same in the SU(2)$\times$SU(2) case and in the SU(3)$\times$ SU(3) case (see Sec.~19.7 of ~\cite{Weinberg2})}
\be \tau \sim 30 f^3 \label{kappaf}. \ee
This relation should also be true in this version of QCD with $f$ taken at the Planck scale as long as the elements of $M_q$ are  much smaller than $f$.

 Now, by choosing a hierarchy $m_{\tilde  u} \gg m_{\tilde d}, m_{\tilde  s}$ (which is inverted compared to the observed quark-mass case) one obtains that the lightest pseudo-Goldstone boson is the complex scalar $\tilde K^0$.
During inflation  we can parameterize it as 
\be\tilde K^0 = \frac{\phi}{\sqrt{2}} \exp(i\delta/f),\ee where $\phi$ is the real inflaton field and $\delta$ is some angular real field. To compute the low energy potential for $\tilde K^0$ one can  set all remaining (very heavy) tilde-meson fields to zero in~(\ref{Bf}):
 \be \label{Bf2}   B= \left(\baccc 0& 0 &0 \\
0 & 0 &\frac{\phi}{\sqrt{2}}  \,  e^{i\delta/f} \\
0 &\frac{\phi}{\sqrt{2}}  \, e^{-i\delta/f} & 0 \ea \right). 
\ee
In this case the eigenvalues of $B$ are $0$,  $\phi/\sqrt{2}$ and $-\phi/\sqrt{2}$ so
 \be \label{cBf}  \cos\left(\frac{\sqrt{2}B}{f}\right)= P +\cos\left(\frac{\phi}{f}\right) (1-P),
\ee
where $P$ is the projector on the first (vanishing) eigenvalue of $B$ (i.e.~$P$ = diag$(1,0,0)$). 
From~(\ref{VNq}) the potential reads
 \be V(\phi) = \tau (m_{\tilde d}+m_{\tilde s}) \cos\left(\frac{\phi}{f}\right) +\Lambda_0+\tau m_{\tilde u}\ee
or, equivalently,
  \be V(\phi) = \Lambda^4 \left(1+\cos\left(\frac{\phi}{f}\right)\right) +\Lambda_{\rm cc}, \label{VInf}\ee
having identified   $\Lambda = [\tau (m_{\tilde d}+m_{\tilde s})]^{1/4}$  and  $\Lambda_0 = \Lambda^4+\Lambda_{\rm cc}-\tau m_{\tilde u}$. Therefore, the mass scale $\Lambda$ is sourced by the symmetry breaking parameters in the fundamental action (in this case the tilde-quark masses): one has $\Lambda\to 0$ as $m_{\tilde d}+m_{\tilde s} \to 0$. Since $V$ is independent of $\delta$,  slow-roll inflation occurs along trajectories with constant $\delta$  (the slow-roll approximation is used to analyze also this inflationary scenario).
 
 The equations derived so far allow us to estimate the masses  of the lightest tilde-quarks and  that of the lightest tilde-meson, $m_\phi$. First note that from the expression in~(\ref{VInf}) one finds $m_\phi=\Lambda^2/f$, where $m_\phi^2$ has been identified with the second derivative of $V$ computed at its point of minimum, namely at $\phi=\pi f$. When $\phi$ gives a dominant contribution to inflation (which occurs when the $\phi$ direction in the scalar-field space is flatter than the other field directions)  the parameter $\Lambda$ is  below the Planck scale, around $10^{-2} \bp$ or $10^{-3} \bp$,
  as a consequence of some constraints coming  from CMB observations (Ade (2016) and Akrami (2020)). Using $m_\phi=\Lambda^2/f$ and CMB constraints on $f$ (again from Ade (2016) and Akrami (2020)) 
  one obtains that $m_\phi$ is around the scale $10^{-5} \bp$. The relations in~(\ref{K0}) and~(\ref{kappaf}) then tell us (barring large hierarchies between $m_{\tilde d}$ and $m_{\tilde s}$) that the masses of the lightest tilde-quarks are around $10^{-13} \bp\sim 10^5~$GeV.  
  
  While $n_s$, computed in a field value corresponding to a large-enough period of inflation, is in agreement with observations,  the ratio $r$ violates a CMB bound, Ade (2021). Like for Planckion and some Higgs(-like) inflationary scenarios, in Sec.~\ref{A mixed recipe: multifield inflation} it is shown how including quadratic-in-curvature terms in the action makes also PNGB inflation compatible with observations\footnote{Alternatively, PNGB inflation can be made compatible with observations in other modified-gravity scenarios, which are not covered here (see Ferreira (2018), Simeon (2020), Salvio (2024a), Racioppi (2024) and also Pradisi (2022) for some background material).}.

\section{A mixed recipe: multifield inflation}\label{A mixed recipe: multifield inflation}

We can combine all the single-field inflaton scenarios discussed so far to obtain a multifield inflation. Higgs(-like) inflation and Planckion inflation are clearly particular cases of the general theory discussed in Subsec.~\ref{Scalaron inflation}. But even PNGB inflation illustrated in Sec.~\ref{Composite inflaton: the Goldstone case} can be considered a particular case of that general theory: first, as we have seen, PNGB inflation may be driven by a PNGB arising from a gauge theory similar to QCD (but with a confining scale $f$ at a much higher energy) and such a theory is included in the  general theory of Subsec.~\ref{Scalaron inflation}; second, effectively, at energy below $f$ the PNGB potential in~(\ref{VInf}) can be included in the full potential by extending the definition in~(\ref{RenPot}), but without affecting the validity of~(\ref{LE}) and~(\ref{Udef}). 

As we have seen Planckion inflation and PNGB inflation, as well as some forms of Higgs(-like) inflation, suffer from a too large value of $r$, which exceeds a CMB constraint. This problem can be circumvented in the multifield case: scalaron inflation and Higgs inflation (for  a large value of $\xi$) are in excellent agreement with data, so if such scalaron or Higgs directions are flatter enough than the other directions in the scalar-field space we are guaranteed to have agreement with data.

Alternatively, we can include the Weyl-squared term with a large-enough coefficient (a small enough $f_2$) such that it becomes relevant during inflation. This renders the theory perturbatively renormalizable, Stelle (1977), because the Lagrangian is now the most general  one compatible with high-energy scale invariance, barring covariant divergences in $\mathscr{L}$, as clear from~(\ref{Lquad3}). In a perturbatively renormalizable  theory the growth of gravitational interactions as a function of the energy stops at some energy scale $\Lambda_G$ (related to the coefficients of the quadratic-in-curvature terms) above which gravitational interactions start to decrease. Indeed, if this were not the case, the theory would not remain perturbative. So when $\Lambda_G$ is sufficiently small compared to the typical energy scales of inflation the interaction of the graviton gets suppressed and so does $r$, (Salvio (2017,2022), Anselmi (2020), Dondarini (2023)). For this and other reasons quadratic-in-curvature terms have attracted much interest recently and in some works (see Salvio (2024b) and references therein) it was shown that renormalizability can be achieved in a consistent quantum theory.

%\begin{BoxTypeA}[sec5:box1]{This box has a title}
%	\section*{We can have a section in the box}
%	Some text, some text, some text, some text, some text.
%	
%	\subsection*{And even a subsection}
%	Some text, some text, some text, some text, some text.
%\end{BoxTypeA}

%%%%%%%%%%%%%%%%%%%%%%%%%%%%%%%%%%%%%%%%%
%% Mandatory: A concluding paragraph summing up your main points in the chapter
%% Optional: Also include big questions in the field that are still to be answered. What topics/methods/questions are researchers like to focus on next?
\section{Conclusions}
\label{sec:conclusions}
%Mandatory: A concluding paragraph summing up the main points of your chapter.\\
%Optional: Also include big questions in the field that are still be be answered. 
%What topics/methods/questions are our colleagues like to focus on next?
Let us conclude this chapter by giving a detailed summary of the main points addressed. The main purpose of this chapter was to explain at an introductory level how well-established particle-physics models and/or mechanisms can naturally lead to inflation and how this can provide testable predictions that can help us find new physics effects.

\vspace{0.2cm}

To achieve this goal the following two well-established features of theoretical particle physics have been shown to provide an essential characteristic of inflation, a naturally flat potential.
\begin{enumerate}
\item The first feature is scale invariance which occurs, at least classically, in the high-energy regime of microscopic particle-physics theories. This is because at high energies terms in the action with dimensionful coefficients can be neglected and renormalizability implies that one is left with dimensionless coefficients only. The scale invariance of such high-energy theory is typically broken  by quantum corrections, which are, however, small if perturbation theory holds. This results in a quasi-flat potential which is suitable for inflation and ensures a graceful exit from it. In this context the inflatons are interpreted as elementary particles because they are field variables in microscopic QFTs. The above-mentioned scale invariance can be extended to the full theory, including the gravity sector, meaning that one can radiatively generate the Higgs VEV, the Planck scale and the cosmological constant through DT, unlike in the SM. This has the advantage of explaining large hierarchies between mass scales, such as the observed hierarchy $v_h\ll M_P$, because the running of the renormalized parameters is only logarithmic in $\mu$. 
\item The second feature is Goldstone's theorem, which characterizes the physics of spontaneously broken global symmetries: the potential of a Nambu-Goldstone boson is flat, but acquires some steepness if explicit symmetry-breaking terms are introduced. Such PNGB is, therefore, a natural BSM inflaton candidate. An explicit QCD-like completion of such scenario has been illustrated in detail. 
\end{enumerate}

\vspace{0.2cm}

Furthermore, a general particle-physics motivated theory of inflation, which combines the above-mentioned features, has been constructed in Subsec.~\ref
{Scalaron inflation} and Sec.~\ref{A mixed recipe: multifield inflation}. 

\vspace{0.2cm}

The future of inflation is bright, future observational activities, such as LiteBIRD, are planned to study in greater detail the CMB to discriminate between various inflationary models. So, these particle-physics and BSM ideas will be soon tested.

\begin{ack}[Acknowledgments]%
 I thank all researchers that have collaborated with me on the topic of this chapter.
\end{ack}

%%%%%%%%%%%%%%%%%%%%%%%%%%%%%%%%%%%%%%%%%%%%
%% Optional: A list of references to other relevant works/articles/websites which are not cited in the text but that would further enhance a readers understanding of this topic
%\seealso{article title article title}

%%%%%%%%%%%%%%%%%%%%%%%%%%%%%%%%%%%%%%%%%
%% Mandatory: Bibliography using bibtex 
\bibliographystyle{Numbered-Style} %% for Numbered Reference Style

\bibliography{reference}

\end{document}